\documentclass[conference]{IEEEtran}

\usepackage{amsmath,amssymb,amsfonts}
\usepackage{graphicx}
\usepackage{textcomp}
\usepackage{xcolor}
\usepackage{algorithm}
\usepackage{algpseudocode}
\usepackage{amsmath}

\title{DroneXNFT: An NFT-Driven Framework for Secure Autonomous UAV Operations and Flight Data Management}

\author{
    \IEEEauthorblockN{Khaoula Hidawi~\IEEEmembership{Member,~IEEE}}
    \IEEEauthorblockA{
        Department of Theoretical and Applied Science (DISTA),\\
        Università degli Studi dell'Insubria, Varese, Italy.\\
        khidawi@uninsubria.it
    }
}

\begin{document}

\maketitle

\begin{abstract}
Non-Fungible Tokens (NFTs) have emerged as a revolutionary method for managing digital assets, providing transparency and secure ownership records on a blockchain. In this paper, we present a theoretical framework for leveraging NFTs to manage UAV (Unmanned Aerial Vehicle) flight data. Our approach focuses on ensuring data integrity, ownership transfer, and secure data sharing among stakeholders. This framework utilizes cryptographic methods, smart contracts, and access control mechanisms to enable a tamper-proof and privacy-preserving management system for UAV flight data.
\end{abstract}

\begin{IEEEkeywords}
NFTs, Blockchain, UAV Flight Data, Smart Contracts, Cryptography, Data Integrity, Secure Data Sharing
\end{IEEEkeywords}

\section{Introduction}

The exponential growth of blockchain technology and the rise of Non-Fungible Tokens (NFTs) have transformed the way digital assets are managed and exchanged. Originally popularized in the context of digital art and collectibles, NFTs have demonstrated their potential for broader applications due to their unique properties of ownership verification, tamper resistance, and traceability on decentralized networks \cite{prihatno2023}. Blockchain technology, with its immutability and distributed consensus, provides a secure and transparent medium for recording and verifying ownership \cite{10411476, rawat2023}. Together, blockchain and NFTs create a robust framework for managing the authenticity and provenance of digital assets in various domains, including data management \cite{sangarapillai2023}, identity \cite{HAWASHIN2024100597}, and beyond \cite{prihatno2023}.

In the context of UAV (Unmanned Aerial Vehicle) flight operations, vast amounts of data are generated during missions. This data includes critical flight parameters such as three-dimensional flight paths, sensor readings, timestamps, and environmental data, all of which are valuable for purposes such as surveillance, logistics, disaster management, and scientific research. Ensuring the integrity and authenticity of UAV flight data is crucial, particularly when the data is shared between multiple stakeholders or used for decision-making \cite{pungila2022}. UAV data may also involve sensitive information that necessitates secure handling, including protecting against unauthorized access, tampering, or corruption.
Traditionally, UAV flight data is stored in centralized systems, which may expose it to risks such as data breaches, tampering, or loss. Centralized systems also face challenges related to transparent ownership, access control, and verifying the integrity of the data. In such environments, ensuring that the data has not been altered, and proving its provenance, can be complex and resource-intensive. This paper proposes a decentralized, blockchain-based solution to address these challenges by leveraging NFTs to encapsulate UAV flight data as a tamper-proof, traceable digital asset \cite{CH2020102670}.

NFTs, which are unique, non-interchangeable tokens stored on a blockchain, allow for the creation of a digital representation of the UAV flight data that is immutable and easily verifiable . Each NFT can represent a specific set of flight data, providing clear ownership and a verified history of how the data has been handled. By storing critical metadata, such as the UAV’s flight path, sensor readings, and timestamps, within the NFT, the authenticity of the data can be guaranteed across its lifecycle. The decentralized nature of blockchain ensures that once data is recorded and minted into an NFT, it cannot be altered without invalidating the cryptographic proofs that secure it \cite{10152611, toumi2021}.
Furthermore, the integration of smart contracts into this system enables automated, rule-based interactions between parties. Smart contracts are self-executing pieces of code that run on a blockchain and enforce predefined rules \cite{wang2020}. In the context of UAV data management, smart contracts can be used to regulate data access, ownership transfers, and licensing agreements in a transparent and automated manner \cite{518607}. This reduces the need for intermediaries and ensures that stakeholders adhere to the conditions under which data can be accessed or transferred.

By employing cryptographic techniques, such as hashing and encryption, our framework also enhances data security \cite{mattsson2021}. Cryptographic hash functions ensure the integrity of the data, while encryption mechanisms provide privacy and access control, ensuring that only authorized users can view or modify the data. This is particularly important for sensitive UAV missions, where data confidentiality must be preserved while enabling secure collaboration between parties.

In this paper, we propose a blockchain-based framework for managing UAV flight data using NFTs. Our framework addresses several key challenges, including the secure representation of flight data, transparent ownership and transfer of data, and controlled access through smart contracts. The use of NFTs for flight data management introduces a novel approach to data authentication and provenance, while the blockchain ensures that all interactions with the data are secure, traceable, and tamper-resistant.The use of blockchain and NFTs in UAV data management offers new possibilities for ensuring data traceability and ownership. Recent research \cite{9395816,wang2020, tychola2024} has shown that blockchain-based traceability systems can efficiently certify UAV components, ensuring transparency in the entire lifecycle of UAV operations \cite{HAWASHIN2024100597}. By incorporating NFTs, the traceability of individual UAV parts and the flight data itself becomes decentralized and verifiable, enabling secure exchanges of ownership without relying on centralized authorities.

\begin{figure}[H]
    \centering
    \includegraphics[width=\linewidth, height = 9cm]{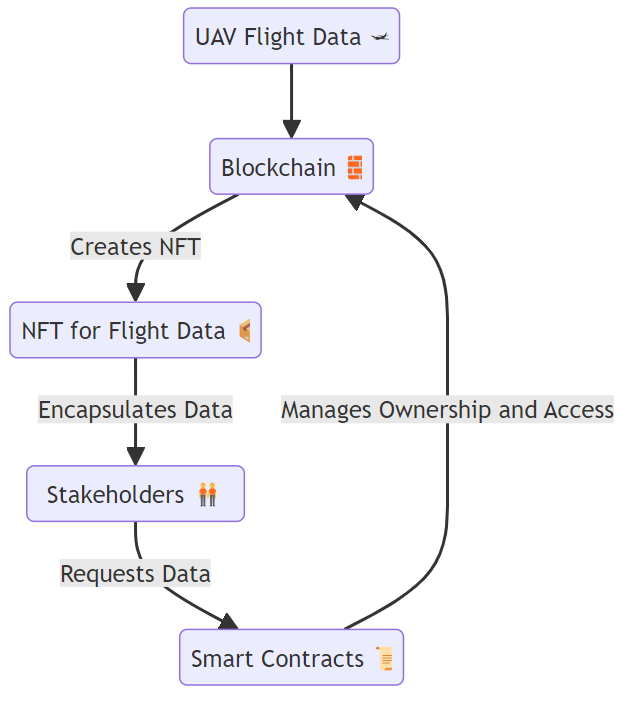}
    \caption{System Architecture for UAV Flight Data Management Using NFTs and Blockchain}
    \label{fig:enter-label}
\end{figure}

The primary contributions of this paper are:
\begin{itemize}
    \item A blockchain-based system for securely managing UAV flight data using NFTs.
    \item A method for ensuring the authenticity and integrity of UAV data through cryptographic techniques.
    \item The use of smart contracts for secure data sharing, ownership transfer, and automated compliance with access control policies.
    \item Consideration of privacy and security concerns in the handling of sensitive UAV data.
\end{itemize}

This theoretical framework sets the foundation for a decentralized and secure approach to UAV data management, with applications extending to real-world scenarios where trust and data integrity are of paramount importance.

\section{Blockchain-Based Authentication and Data Integrity}

\subsection{Cryptographic Hash Functions}
Ensuring the integrity of UAV flight data is paramount. To achieve this, we use cryptographic hash functions. A hash function \(H(x)\) is defined as:

\begin{equation}
H(x) = h, \quad x \in \{0,1\}^n, \ h \in \{0,1\}^{256}
\end{equation}

where \(x\) represents the UAV flight data and \(h\) is the corresponding fixed-length hash value. This ensures that even a small change in the input data will produce a completely different hash, making any tampering evident.

\subsection{Merkle Trees for Efficient Verification}
To efficiently verify large UAV flight datasets, we utilize Merkle trees \cite{9588047}. Each leaf node of the Merkle tree represents a hash of a data block (e.g., flight data at a specific timestamp), and the root of the Merkle tree provides a hash representing the entire dataset. 

Given a dataset \(D = \{d_1, d_2, \dots, d_n\}\), the root hash is:

\begin{equation}
H_{\text{root}} = H(H(d_1, d_2), H(d_3, d_4), \dots, H(d_{n-1}, d_n))
\end{equation}

The Merkle tree allows any part of the dataset to be efficiently authenticated without needing to rehash the entire dataset.

\begin{figure}[H]
    \centering
    \includegraphics[width=\linewidth, height = 4cm]{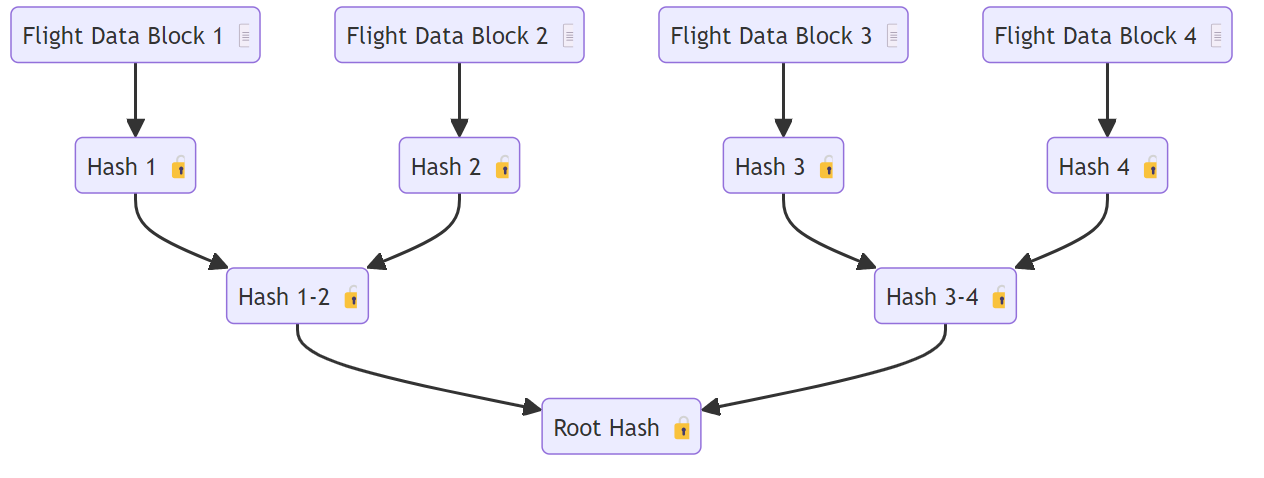}
    \caption{Merkle Tree for Efficient UAV Flight Data Verification}
    \label{fig:enter-label}
\end{figure}

\section{NFTs for Data Ownership and Provenance}

\subsection{NFT Minting Process}
NFTs are used as digital certificates of ownership for UAV flight data. Each NFT encapsulates a unique reference to the flight data and associated metadata. The minting process for an NFT representing flight data is as follows:

\begin{equation}
\text{NFT}(D) = \{ \text{Token ID}, H(D), \text{metadata} \}
\end{equation}

Here, \(H(D)\) is the hash of the UAV flight dataset \(D\), ensuring that the NFT represents the authentic and verified data. The metadata includes mission details, timestamps, and other critical information, all registered on the blockchain.

\begin{figure}[H]
    \centering
    \includegraphics[width=6cm, height = 7cm]{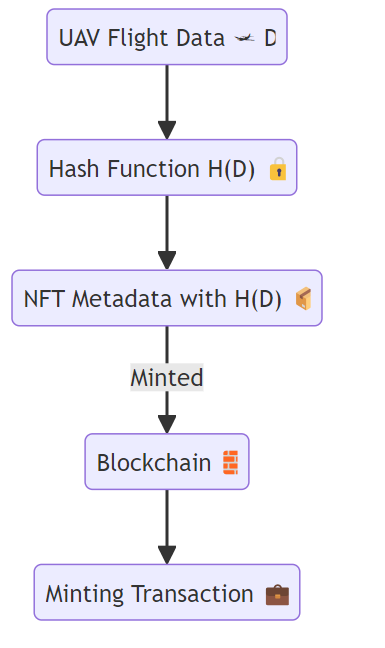}
    \caption{NFT Minting Process for UAV Flight Data}
    \label{fig:enter-label}
\end{figure}

\subsection{Ownership Transfers}



Ownership of an NFT can be transferred securely between parties using a blockchain transaction. The transfer process is governed by a smart contract, ensuring that only the authorized owner of the NFT can execute transfers. The following algorithm details the smart contract function used for transferring NFTs.

\begin{algorithm}[H]
\caption{NFT Ownership Transfer}
\begin{algorithmic}[1]
\State \textbf{Input:} $A_{\text{from}}$: Address of current owner, $A_{\text{to}}$: Address of recipient, $tokenId$: Unique identifier of the NFT
\State \textbf{Output:} Updated ownership of $tokenId$ transferred to $A_{\text{to}}$

\Procedure{TransferNFT}{$A_{\text{from}}, A_{\text{to}}, tokenId$}
    \State \textbf{Require:} $msg.sender == OwnerOf(tokenId)$ \Comment{Verify that the sender is the current owner}
    \If{$msg.sender$ is the owner of $tokenId$}
        \State Execute the transfer of $tokenId$ from $A_{\text{from}}$ to $A_{\text{to}}$
        \State \Call{RecordTransaction}{$A_{\text{from}}, A_{\text{to}}, tokenId$} \Comment{Log transfer on the blockchain}
        \State Update the owner of $tokenId$ to $A_{\text{to}}$
    \Else
        \State \textbf{Revert:} "Only the owner can transfer"
    \EndIf
\EndProcedure
\end{algorithmic}
\end{algorithm}


The ownership transfer algorithm ensures the secure transfer of an NFT between two parties using a blockchain-based smart contract. The function takes three inputs: the current owner's address ($A_{\text{from}}$), the recipient's address ($A_{\text{to}}$), and the unique identifier of the NFT ($tokenId$). The process begins by verifying that the sender of the transaction ($msg.sender$) is indeed the current owner of the NFT. This verification step is crucial to prevent unauthorized transfers. If the check passes, the function proceeds to execute the transfer, updating the ownership of the NFT to the recipient and logging the transaction immutably on the blockchain through a call to the \texttt{RecordTransaction} function. This ensures that the transfer is transparent and can be publicly verified, adding an immutable entry to the blockchain's record of ownership. If the sender is not the current owner, the function is reverted, and the transfer is prevented, ensuring that only the rightful owner can initiate the transfer. By incorporating these checks and immutability through blockchain logging, the algorithm guarantees that NFT transfers are secure, traceable, and irreversible, thereby preserving the integrity of the ownership system.

\begin{figure}[H]
    \centering
    \includegraphics[width=5cm, height = 7cm]{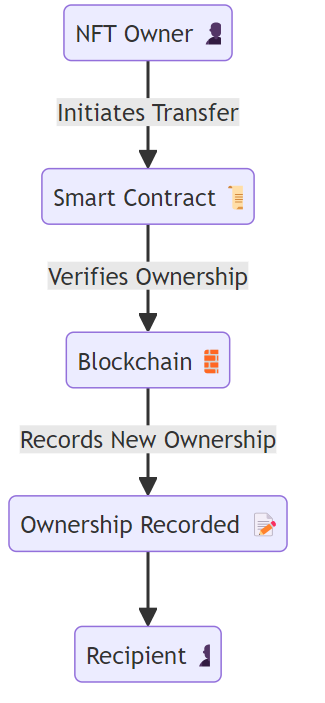}
    \caption{Ownership Transfer Process for UAV Data NFTs}
    \label{fig:enter-label}
\end{figure}

\section{Secure Data Sharing via Smart Contracts}

One of the critical challenges in managing UAV flight data is ensuring that sensitive information is securely shared among stakeholders while maintaining control over who can access the data and under what conditions. Our approach leverages smart contracts to implement access control and licensing mechanisms in a decentralized manner. Additionally, we propose using cryptographic tools like Zero-Knowledge Proofs (ZKPs) to ensure that sensitive data can be verified without exposing the actual contents, ensuring privacy while maintaining verifiability.

\subsection{Access Control and Licensing}
In traditional systems, access control is often implemented through centralized mechanisms, which can be vulnerable to single points of failure or tampering \cite{10071593}. By shifting to a decentralized blockchain-based approach, we ensure that access control is enforced through immutable smart contracts, providing transparency and security.

In our proposed framework, UAV flight data is encapsulated in NFTs, with each NFT corresponding to a specific flight dataset \( D \). The ownership of each NFT is tracked on the blockchain, and the data can only be accessed or shared by authorized parties through predefined conditions encoded in smart contracts. Specifically, the smart contract governs licensing agreements, granting or revoking access to the data based on the owner's permissions.

\begin{figure}[H]
    \centering
    \includegraphics[width=7cm, height = 8cm]{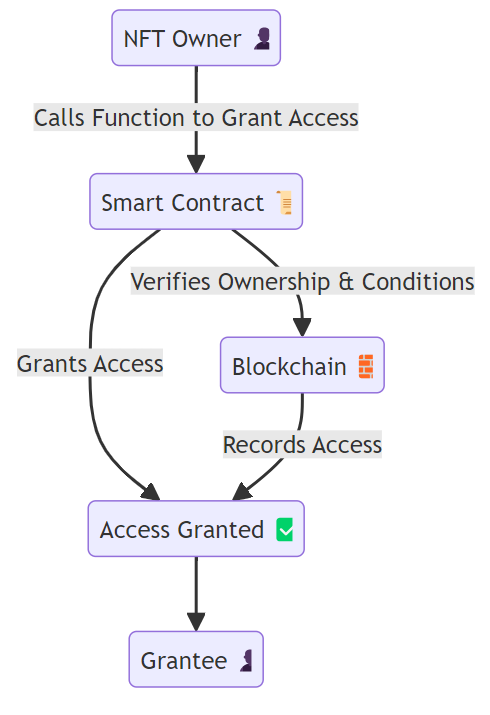}
    \caption{Access Control and Licensing Mechanism for UAV Data Sharing}
    \label{fig:enter-label}
\end{figure}

\subsubsection{Access Control Algorithm}
The algorithm for granting access to an NFT containing UAV flight data can be defined as follows:

\begin{itemize}
    \item \textbf{Input:}
    \begin{itemize}
        \item \( A_{\text{owner}} \): Address of the current NFT owner (data owner).
        \item \( A_{\text{grantee}} \): Address of the party to whom access is to be granted (data grantee).
        \item \( \text{tokenId} \): The unique identifier of the NFT corresponding to the flight data.
        \item \( T_{\text{expiration}} \): Expiration timestamp for the access grant.
        \item \( P_{\text{contract}} \): The predefined conditions for access stored in the smart contract (e.g., access fees, time limits).
    \end{itemize}
\end{itemize}

\subsubsection{Access Control Smart Contract Function}


The following algorithm represents the smart contract function to grant access to a specific NFT based on predefined conditions. The function ensures that only the current owner of the NFT can grant access to other users, and the access is limited by a specified expiration time.

\begin{algorithm}[H]
\caption{Grant Access to UAV Flight Data NFT}
\begin{algorithmic}[1]
\Require $A_{\text{grantee}}$, $tokenId$, $T_{\text{expiration}}$
\State \textbf{Input:} Address of grantee $A_{\text{grantee}}$, token ID $tokenId$, expiration time $T_{\text{expiration}}$
\State \textbf{Output:} Updated access permissions for $A_{\text{grantee}}$
\Procedure{GrantAccess}{$A_{\text{grantee}}, tokenId, T_{\text{expiration}}$}
    \State \textbf{Require:} $msg.sender == OwnerOf(tokenId)$
    \If{$msg.sender$ is the owner of $tokenId$}
        \State Grant access to $A_{\text{grantee}}$ for $tokenId$
        \State Set access expiration time to $T_{\text{expiration}}$
        \State Store $accessPermissions[A_{\text{grantee}}][tokenId] \gets T_{\text{expiration}}$
    \Else
        \State \textbf{Revert:} "Only the owner can grant access"
    \EndIf
\EndProcedure
\end{algorithmic}
\end{algorithm}

\noindent
The process flow is as follows:
\begin{enumerate}
    \item The data owner \( A_{\text{owner}} \) calls the \texttt{grantAccess} function on the smart contract.
    \item The smart contract verifies that \( A_{\text{owner}} \) is the legitimate owner of the NFT corresponding to \texttt{tokenId}.
    \item The smart contract records access permissions, allowing \( A_{\text{grantee}} \) access to the data until the specified expiration time \( T_{\text{expiration}} \).
    \item When \( A_{\text{grantee}} \) attempts to access the data, the smart contract verifies that the current time \( t_{\text{now}} \) is less than \( T_{\text{expiration}} \) before granting access.
\end{enumerate}

This ensures that access is time-limited and can be revoked automatically after \( T_{\text{expiration}} \), enforcing predefined licensing agreements without the need for manual intervention. The algorithm ensures that only the current owner of the NFT can grant access.

\subsubsection{Mathematical Model for Licensing}

Let \( D \) be the dataset corresponding to the flight data encapsulated in the NFT. Access to the dataset \( D \) is only granted if the following condition holds:

\begin{equation}
t_{\text{now}} < T_{\text{expiration}} \quad \text{and} \quad P_{\text{contract}} = \text{True}
\end{equation}

Where:
\begin{itemize}
    \item \( t_{\text{now}} \) is the current timestamp.
    \item \( T_{\text{expiration}} \) is the expiration timestamp for access.
    \item \( P_{\text{contract}} \) represents the conditions set in the smart contract, which could include factors such as required payments, geographic restrictions, or usage limitations.
\end{itemize}

If the conditions are met, the smart contract grants the grantee access to the data. This mechanism ensures that all parties adhere to the predefined conditions, with enforcement handled programmatically by the blockchain.

\subsection{Zero-Knowledge Proofs for Data Verification}
In addition to controlling access to the data, it is often necessary for one party to prove that they possess valid UAV flight data without revealing the actual contents of the data. This is especially important when the data is sensitive, and the verifier should only confirm the data's integrity without accessing the underlying details. Zero-Knowledge Proofs (ZKPs) \cite{10293092} provide a powerful cryptographic tool for achieving this.

\subsubsection{Zero-Knowledge Proof Algorithm}

We propose utilizing Zero-Knowledge Succinct Non-Interactive Arguments of Knowledge (zk-SNARKs) for proving the correctness of UAV flight data \cite{10616301}. zk-SNARKs allow a prover to convince a verifier that they possess a valid piece of data (in this case, the flight data \( D \)) without revealing the data itself.

\begin{figure}[H]
    \centering
    \includegraphics[width=7cm, height = 8cm]{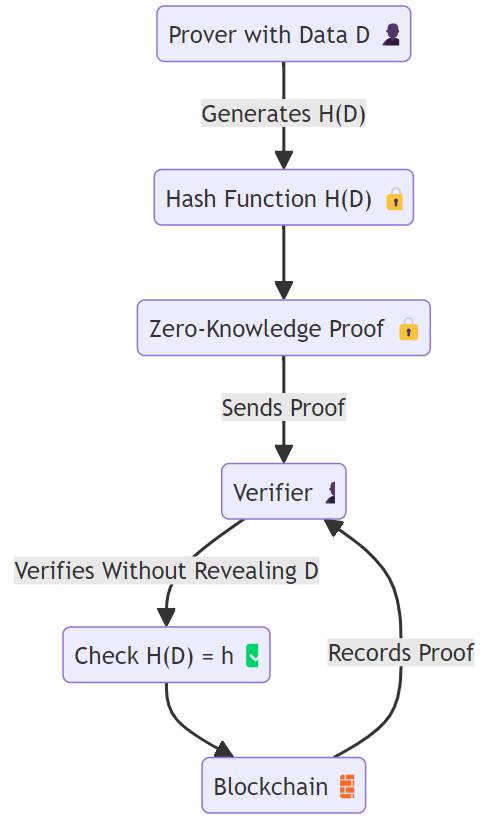}
    \caption{Zero-Knowledge Proofs for UAV Data Integrity Verification}
    \label{fig:enter-label}
\end{figure}

The prover \( P \) wishes to prove to a verifier \( V \) that they possess the correct flight data \( D \), which corresponds to the hash \( H(D) \) stored in the NFT metadata.

\begin{itemize}
    \item \textbf{Input:}
    \begin{itemize}
        \item \( D \): The flight data (known to the prover but hidden from the verifier).
        \item \( H(D) \): The cryptographic hash of the flight data stored on the blockchain.
        \item zk-SNARK proof \( \pi \).
    \end{itemize}
\end{itemize}

\noindent
The zk-SNARK system consists of the following phases:
\begin{itemize}
    \item \textbf{Setup:} A trusted setup generates public parameters \( pp \).
    \item \textbf{Proving:} The prover \( P \) generates a proof \( \pi \) that proves the existence of \( D \) such that:
    \begin{equation}
    \text{ZKP}\{ \exists D : H(D) = h \}
    \end{equation}
    without revealing \( D \) itself. This is done by constructing a zk-SNARK proof \( \pi \) using the public parameters \( pp \).
    \item \textbf{Verification:} The verifier \( V \) checks the validity of the proof \( \pi \) without learning the contents of \( D \). The verification equation is:
    \begin{equation}
    V(\pi, H(D), pp) = \text{True}
    \end{equation}
\end{itemize}

If the proof is valid, the verifier is convinced that \( P \) possesses the valid data corresponding to \( H(D) \), without accessing the actual data.

\subsubsection{Mathematical Representation of ZKP}

Let \( H(D) \) be the hash of the dataset \( D \), and \( \mathcal{P} \) be the statement that needs to be proven:
\begin{equation}
\mathcal{P} = \{ D : H(D) = h \}
\end{equation}

The prover constructs the proof \( \pi \) that demonstrates the existence of \( D \) such that the statement \( \mathcal{P} \) holds, without revealing \( D \). The verifier then uses the public verification key \( vk \) to check the proof:
\begin{equation}
V(vk, \pi) = \text{True}
\end{equation}

This ensures that the prover holds the valid flight data while maintaining the confidentiality of the data.

\begin{figure}[H]
    \centering
    \includegraphics[width=10.5cm, height = 8cm]{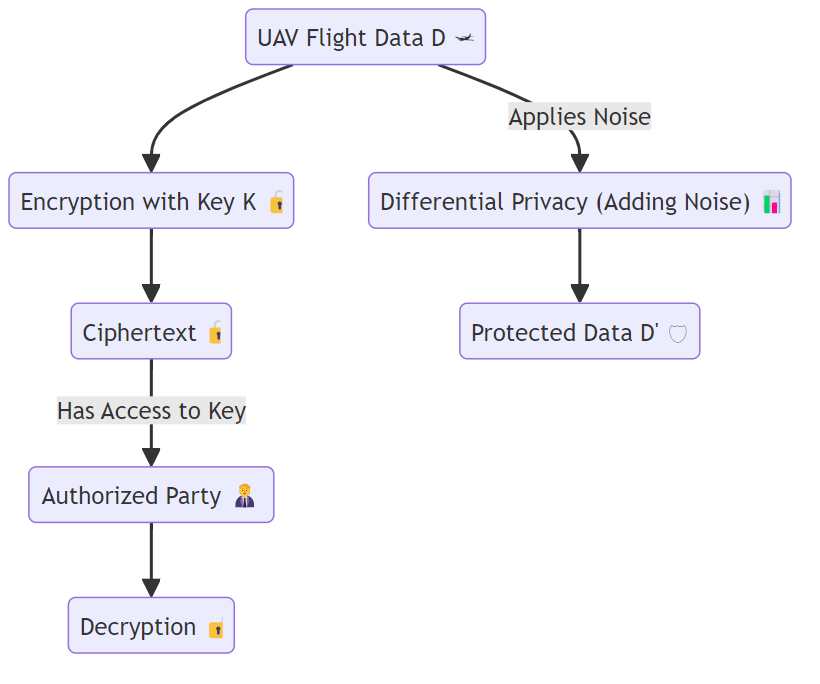}
    \caption{Data Encryption and Differential Privacy for UAV Data Protection}
    \label{fig:enter-label}
\end{figure}

\subsubsection{Application in UAV Data}

In the context of UAV flight data, ZKPs can be used in scenarios where a stakeholder needs to verify the authenticity of data without revealing sensitive information, such as geographic coordinates or sensor readings. For example, in collaborative missions, one organization might need to prove that their UAV followed a specified flight path without revealing the exact locations. ZKPs allow such proofs to be made securely, ensuring both privacy and verifiability.


\section{Decentralized Autonomous UAV Operations (DAO-UAV)}

To further enhance the capabilities of UAV data management and extend the potential of our NFT-driven framework, we propose a novel extension: Decentralized Autonomous UAV Operations (DAO-UAV). This concept leverages blockchain, NFTs, and smart contracts to create a fully autonomous, decentralized system for managing UAV fleets. The DAO-UAV model introduces a new paradigm where UAV operations, data collection, task assignment, and decision-making processes are managed without the need for centralized control. This section details the mathematical models and algorithms underpinning the DAO-UAV framework, including task assignment and ownership transfer.

\subsection{Autonomous UAV Task Assignment}

In the DAO-UAV framework, each UAV is represented as an NFT on the blockchain, encapsulating its operational metadata, ownership, and performance history. Smart contracts are employed to automate the assignment of tasks to UAVs based on predefined criteria such as location, payload capacity, and mission urgency. The smart contract autonomously assigns tasks by checking the blockchain for available UAVs that meet these task-specific requirements.


Let \( U = \{U_1, U_2, \dots, U_n\} \) represent the set of available UAVs, and let a task \( T \) have the following requirements:
\[
T = \{L_t, P_t, M_t\}
\]
where:
\begin{itemize}
    \item \( L_t \) denotes the task's location,
    \item \( P_t \) represents the required payload capacity,
    \item \( M_t \) indicates the mission urgency.
\end{itemize}

Each UAV \( U_i \in U \) is characterized by:
\[
U_i = \{L_i, P_i, S_i\}
\]
where:
\begin{itemize}
    \item \( L_i \) is the UAV's current location,
    \item \( P_i \) is the payload capacity of the UAV,
    \item \( S_i \) is the UAV's status (e.g., available, in mission).
\end{itemize}

The task is assigned to UAV \( U_i \) if the following conditions hold:
\[
U_i \ \text{is assigned to} \ T \ \text{if:}
\quad S_i = \text{available}, \ L_i \approx L_t, \ \text{and} \ P_i \geq P_t
\]
Here, the UAV with the smallest distance to the task location \( L_t \) is chosen, provided it meets the required payload capacity \( P_t \).


\begin{algorithm}[H]
\caption{Task Assignment via Smart Contract}
\begin{algorithmic}[1]
\State \textbf{Input:} Task \( T = \{L_t, P_t, M_t\} \), UAV set \( U = \{U_1, U_2, \dots, U_n\} \)
\State \textbf{Output:} Selected UAV \( U_{\text{selected}} \)
\Procedure{AssignTask}{$T, U$}
    \State Initialize \( U_{\text{selected}} \gets \text{None} \)
    \State Initialize \( d_{\text{min}} \gets \infty \)
    \ForAll{$U_i \in U$}
        \If{$S_i = \text{available} \ \textbf{and} \ P_i \geq P_t$}
            \State Compute distance $d_i = \|L_i - L_t\|$
            \If{$d_i < d_{\text{min}}$}
                \State $d_{\text{min}} \gets d_i$
                \State $U_{\text{selected}} \gets U_i$
            \EndIf
        \EndIf
    \EndFor
    \State \Return $U_{\text{selected}}$
\EndProcedure
\end{algorithmic}
\end{algorithm}

This algorithm automates task assignments by evaluating available UAVs and selecting the UAV closest to the task location, provided it meets the payload and availability requirements. The use of smart contracts ensures that the task assignment is transparent, secure, and verifiable on the blockchain.

\subsection{Autonomous Ownership and Data Management}

The DAO-UAV framework also facilitates the automated ownership transfer of UAVs and their associated data through smart contracts. When a UAV is transferred to a new operator, its corresponding NFT, containing operational metadata, is transferred securely via the blockchain, ensuring a tamper-proof transaction. This section describes the mathematical model and algorithm for ownership transfer.

\subsubsection{Model for Ownership Transfer}

Let \( O_{\text{current}} \) denote the current owner of UAV \( U_i \), and let \( O_{\text{new}} \) denote the new owner. The transfer process is governed by the smart contract \( SC_{\text{transfer}} \), which ensures the validity of the transaction.

The ownership transfer can be expressed as:
\[
\text{Transfer}(U_i, O_{\text{current}}, O_{\text{new}}) = 
\begin{cases}
\text{Success}, & \text{if} \ SC_{\text{transfer}}(U_i) = \text{valid} \\
\text{Failure}, & \text{otherwise}
\end{cases}
\]
The smart contract verifies the current owner's authenticity and, upon successful validation, transfers ownership and the associated NFT metadata to the new owner, recording the transaction immutably on the blockchain.


\begin{algorithm}[H]
\caption{NFT-Based UAV Ownership Transfer}
\begin{algorithmic}[1]
\State \textbf{Input:} UAV \( U_i \), Current Owner \( O_{\text{current}} \), New Owner \( O_{\text{new}} \)
\State \textbf{Output:} Transfer status
\Procedure{TransferOwnership}{$U_i, O_{\text{current}}, O_{\text{new}}$}
    \If{$SC_{\text{transfer}}(O_{\text{current}}, U_i) = \text{valid}$}
        \State Update ownership record on the blockchain
        \State Transfer associated NFT and metadata to \( O_{\text{new}} \)
        \State \Return \text{Success}
    \Else
        \State \Return \text{Failure}
    \EndIf
\EndProcedure
\end{algorithmic}
\end{algorithm}

This algorithm ensures secure and automated transfers of UAV ownership by verifying the current owner through a smart contract. If the verification succeeds, the NFT representing the UAV is transferred to the new owner, and the blockchain records the transaction immutably.

\subsection{Advantages of DAO-UAV Model}

The DAO-UAV model offers several advantages over traditional centralized UAV management systems, including:

\begin{itemize}
    \item \textbf{Decentralized Decision-Making}: The system eliminates single points of failure by decentralizing task assignment and data management, allowing for more robust and fault-tolerant UAV operations.
    \item \textbf{Autonomous Fleet Management}: UAV fleets can autonomously manage their missions, from task assignment to data recording and validation, using smart contracts that enforce operational rules automatically.
    \item \textbf{Cost and Time Efficiency}: By automating the task distribution and UAV operations, the overhead of manual intervention is minimized, leading to faster and more efficient UAV fleet operations.
\end{itemize}

The integration of NFTs, blockchain, and smart contracts into a decentralized autonomous framework enables UAV fleets to operate with greater transparency, security, and scalability. This novel DAO-UAV approach is ideal for large-scale UAV operations, such as disaster management, logistics, and surveillance, where decentralized control and real-time data validation are critical.

\section{Security and Privacy Considerations}

\subsection{Data Encryption}
Flight data is encrypted to ensure that only authorized parties can access it. The encryption process uses a symmetric key \(K\) as follows:

\begin{equation}
E_K(D) = \text{ciphertext}
\end{equation}

Here, \(E_K\) is a symmetric encryption function (e.g., AES), and only parties possessing the key \(K\) can decrypt the flight data.

\subsection{Differential Privacy}

To protect sensitive information in UAV data, we propose incorporating differential privacy techniques. By adding Gaussian noise to the dataset, we ensure that individual data points cannot be re-identified:

\begin{equation}
D' = D + \mathcal{N}(0, \sigma^2)
\end{equation}

where \(\mathcal{N}(0, \sigma^2)\) is Gaussian noise with variance \(\sigma^2\), ensuring privacy without compromising the utility of the shared data.

\section{Conclusion}
In this paper, we present a theoretical framework for managing UAV flight data using NFTs and blockchain technology. By leveraging cryptographic methods, smart contracts, and access control mechanisms, we ensure the authenticity, ownership, and secure sharing of UAV data. This approach addresses key challenges in data integrity, privacy, and collaboration in UAV operations. Future work will explore real-world applications and further enhance privacy-preserving techniques for UAV data sharing.

\end{document}